\pdfoutput=1
\documentclass[preprint,preprintnumbers,amsmath,amssymb,floatfix,endfloats*]{revtex4}

\usepackage{graphicx}
\usepackage{dcolumn}
\usepackage{bm}
\usepackage{amsfonts}

\DeclareMathAlphabet{\mathsfsl}{OT1}{cmr}{bx}{it}
\begin{document}
\title{Structural relaxation in amorphous materials under cyclic tension-compression loading}
\author{Pritam Kumar Jana$^{1,2}$ and Nikolai V. Priezjev$^{3}$}
\affiliation{$^{1}$Institut f{\"u}r Theoretische Physik,
Universit{\"a}t G{\"o}ttingen, Friedrich-Hund-Platz 1, 37077
G{\"o}ttingen, Germany}
\affiliation{$^{2}$Universite Libre de Bruxelles (ULB),
Interdisciplinary Center for Nonlinear Phenomena and Complex
Systems, Campus Plaine, CP 231, Blvd. du Triomphe, B-1050 Brussels,
Belgium}
\affiliation{$^{3}$Department of Mechanical and Materials
Engineering, Wright State University, Dayton, OH 45435}
%
%
\date{\today}
\begin{abstract}

The process of structural relaxation in disordered solids subjected
to repeated tension-compression loading is studied using molecular
dynamics simulations.  The binary glass is prepared by rapid cooling
well below the glass transition temperature and then periodically
strained at constant volume. We find that the amorphous system is
relocated to progressively lower potential energy states during
hundreds of cycles, and the energy levels become deeper upon
approaching critical strain amplitude from below.  The decrease in
potential energy is associated with collective nonaffine
rearrangements of atoms, and their rescaled probability distribution
becomes independent of the cycle number at sufficiently large time
intervals.   It is also shown that yielding during startup shear
deformation occurs at larger values of the stress overshoot in
samples that were cyclically loaded at higher strain amplitudes.
These results might be useful for mechanical processing of amorphous
alloys in order to reduce their energy and increase chemical
resistivity and resistance to crystallization.

\vskip 0.5in


\end{abstract}

\maketitle

\section{Introduction}

The advancement of processing methods and characterization
techniques is crucial for the rational design of amorphous alloys
with a combination of desired properties, involving strength,
plasticity, corrosion resistance, and wear resistance~\cite{Khan18}.
It is well realized that the exceptionally high strength,
thermoplastic formability, and excellent magnetic properties make
metallic glasses suitable for numerous structural, biomedical, and
magnetic applications~\cite{Khan18}.  In contrast to crystalline
materials, metallic glasses are characterized by an amorphous
structure, and their elementary plastic deformation involves swift
rearrangements of small groups of atoms, or shear
transformations~\cite{Spaepen77,Argon79}.   A major obstacle for the
widespread use of metallic glasses, however, is the strain
localization within narrow bands, and, as a result, a failure of the
material under applied strain~\cite{Egami13}.   Depending on a
technological application, various thermo-mechanical processing
methods might be required to either relax the system to lower energy
states or rejuvenate the glass to higher energies~\cite{Greer16}.
Examples of the most common methods include high-pressure torsion,
wire drawing, shot peening, cryogenic thermal cycling, elastostatic
loading, and cyclic loading~\cite{Greer16}.   However, despite
significant progress, it remains unclear if extreme relaxation or
rejuvenation can be achieved using a combination of these methods;
for example, alternating mechanical agitation and thermal cycling.

\vskip 0.05in

During the last decade, a number of atomistic simulation studies
were carried out to investigate the relaxation dynamics, the range
of accessible energy states, and mechanical properties of
periodically deformed amorphous materials~\cite{
Priezjev13,Sastry13,Reichhardt13,
Priezjev14,Shi15,Yang16,Priezjev16,Kawasaki16,Priezjev16a,Sastry17,
Priezjev17,Zapperi17,Hecke17,Priezjev18,Alava18,Priezjev18a,NVP18strload,She19,
PriMakrho05,PriMakrho09,Deng19,NVP19alt,Priez19ba}. In general, it
was found that the yielding transition occurs after a number of
transient cycles, depending on the preparation history, strain
amplitude and temperature, and, in addition, it is accompanied by
the formation of the system-spanning shear band and a sudden
decrease in the stress amplitude~\cite{Sastry17,Priezjev17,
Priezjev18a,Priez19ba}.   On the other hand, cyclic loading at
strain amplitudes below the critical value can be termed as
`mechanical annealing', which leads to progressively lower potential
energy states over consecutive cycles~\cite{Sastry13,Priezjev18,
Priezjev18a, NVP18strload, NVP19alt}. In the limiting case of
athermal quasistatic periodic loading, amorphous systems eventually
reach the state with the lowest energy, the so-called limit cycle,
where the particle dynamics becomes exactly
reversible~\cite{Sastry13,Reichhardt13}.  Interestingly, it was
recently shown that structural relaxation in amorphous solids is
accelerated when an additional shear orientation is introduced in
the periodic deformation protocol, which leads to the increase in
strength and shear-modulus anisotropy~\cite{NVP19alt}. Nevertheless,
a detailed description of the structural relaxation process during
cyclic loading is still missing.

\vskip 0.05in

In this paper, the influence of time-periodic tension-compression
loading on structural relaxation in amorphous alloys is investigated
using molecular dynamics simulations. The binary alloy is first
rapidly cooled from the liquid state to a temperature well below the
glass transition point, and then periodically strained at constant
volume during hundreds of cycles. It will be shown that cyclic
loading in the elastic range results in lower energy states when
samples are strained at higher strain amplitudes. The relaxation
proceeds via large-scale collective nonaffine displacements of
atoms, and their rescaled probability distribution becomes
independent of the cycle number. The mechanical tests during startup
deformation indicate that the stress overshoot and shear modulus are
increased for samples cyclically loaded at larger strain amplitudes.

\vskip 0.05in

The reminder of the paper is structured as follows. The details of
the simulation model and deformation protocol are described in the
next section.  The time evolution of the potential energy, as well
as the analysis of nonaffine displacements, and the effect of cyclic
loading on mechanical properties are presented in
section\,\ref{sec:Results}. The results are briefly summarized in
the last section.

\section{Molecular dynamics simulations}
\label{sec:MD_Model}


The three-dimensional amorphous material is modeled via the
Kob-Andersen (KA) binary (80:20) Lennard-Jones (LJ)
mixture~\cite{KobAnd95}. In the KA model, the interaction between
atoms of different types is strongly non-additive, which impedes
crystallization at low temperatures~\cite{KobAnd95}. The phase
diagram of the KA mixture is well known, and the model is frequently
employed to study the structural and dynamical properties of glass
formers~\cite{KobAnd95}. More specifically, the LJ interaction
between atoms of types $\alpha,\beta=A,B$ is defined as follows:
\begin{equation}
V_{\alpha\beta}(r)=4\,\varepsilon_{\alpha\beta}\,\Big[\Big(\frac{\sigma_{\alpha\beta}}{r}\Big)^{12}\!-
\Big(\frac{\sigma_{\alpha\beta}}{r}\Big)^{6}\,\Big],
\label{Eq:LJ_KA}
\end{equation}
where $\varepsilon_{AA}=1.0$, $\varepsilon_{AB}=1.5$,
$\varepsilon_{BB}=0.5$, $\sigma_{AA}=1.0$, $\sigma_{AB}=0.8$,
$\sigma_{BB}=0.88$, and $m_{A}=m_{B}$~\cite{KobAnd95}.  All MD
simulations were performed using the LAMMPS parallel code and a
relatively large system of $60\,000$ atoms~\cite{Lammps}.  In order
to reduce the computational time, we use the cutoff radius
$r_{c,\,\alpha\beta}=2.5\,\sigma_{\alpha\beta}$.  Unless noted
otherwise, the physical quantities are expressed in the reduced LJ
units of length, mass, and energy $\sigma=\sigma_{AA}$, $m=m_{A}$,
$\varepsilon=\varepsilon_{AA}$, and, consequently, time
$\tau=\sigma\sqrt{m/\varepsilon}$.   The integration time step in
the velocity-Verlet scheme is $\triangle t_{MD}=0.005\,\tau$.

\vskip 0.05in


The preparation of the poorly annealed glass was performed as
follows. First, the LJ mixture was equilibrated at the high
temperature $T_{LJ}=1.0\,\varepsilon/k_B$ and density
$\rho=\rho_A+\rho_B=1.2\,\sigma^{-3}$.  In the following, the
parameters $k_B$ and $T_{LJ}$ denote the Boltzmann constant and
temperature, respectively.  The system temperature was regulated
through the Nos\'{e}-Hoover thermostat~\cite{Allen87,Lammps}. The
critical temperature of the KA model is $T_c=0.435\,\varepsilon/k_B$
at $\rho=1.2\,\sigma^{-3}$, which was determined by fitting the
diffusion coefficient to a power-law function upon approaching the
glass transition~\cite{KobAnd95}. The linear size of the undeformed
cubic box is $L=36.84\,\sigma$ at the density
$\rho=1.2\,\sigma^{-3}$. Following the equilibration period, the
binary mixture was rapidly cooled to the temperature
$T_{LJ}=0.01\,\varepsilon/k_B$ with the rate
$10^{-2}\varepsilon/k_{B}\tau$, while keeping the volume constant.
This preparation procedure is identical to the one considered in the
recent study on cyclic loading with alternating shear
orientation~\cite{NVP19alt}.

\vskip 0.05in


After the preparation phase, the glass was periodically strained
along the $\hat{z}$ direction as follows:
\begin{equation}
\varepsilon(t)=\varepsilon_0\,\text{sin}(2\pi t/T),
\label{Eq:shear}
\end{equation}
where $\varepsilon_0$ is the strain amplitude and the oscillation
period is $T=5000\,\tau$. The deformation of the simulation cell
involved continuous expansion and contraction of the lateral
dimensions (along the $\hat{x}$ and $\hat{y}$ axes), so that the
total volume remained constant, the so-called pure shear
deformation. The simulations were performed at strain amplitudes,
$\varepsilon_0 \leqslant 0.04$, in the elastic range at
$\rho=1.2\,\sigma^{-3}$ and $T_{LJ}=0.01\,\varepsilon/k_B$. For each
value of the strain amplitude, the glass was subjected to 1400
back-and-forth cycles. During production runs, the potential energy,
stress components, and atomic positions were periodically saved for
further analysis. The data were collected only for one realization
of disorder due to computational constraints.

\section{Results}
\label{sec:Results}


It is well known that dynamic response of disordered solids to
applied mechanical stress or strain depends crucially on the sample
processing history~\cite{Greer16}. In general, after extended time
intervals at temperatures not far below the glass transition
temperature, the amorphous samples become denser and relocate to
lower energy states, whereas rapid cooling from the liquid phase to
low temperatures usually leads to higher energy, less dense
states~\cite{Greer16}. The characteristic feature of mechanical
deformation of relaxed glasses is the appearance of the stress
overshoot at the yielding transition~\cite{Stillinger00}. It was
originally shown that during one subyield shear cycle, the potential
energy landscape becomes tilted, which allows for a collective
rearrangement of groups of atoms to nearby minima, thus leading to a
lower energy state upon strain reversal~\cite{Lacks04}. In the
present study, we explore the relaxation process in binary alloys
during repeated tension-compression deformation and examine the
resulting changes in mechanical properties.

\vskip 0.05in

\begin{figure}[t]
\includegraphics[width=12.0cm,angle=0]{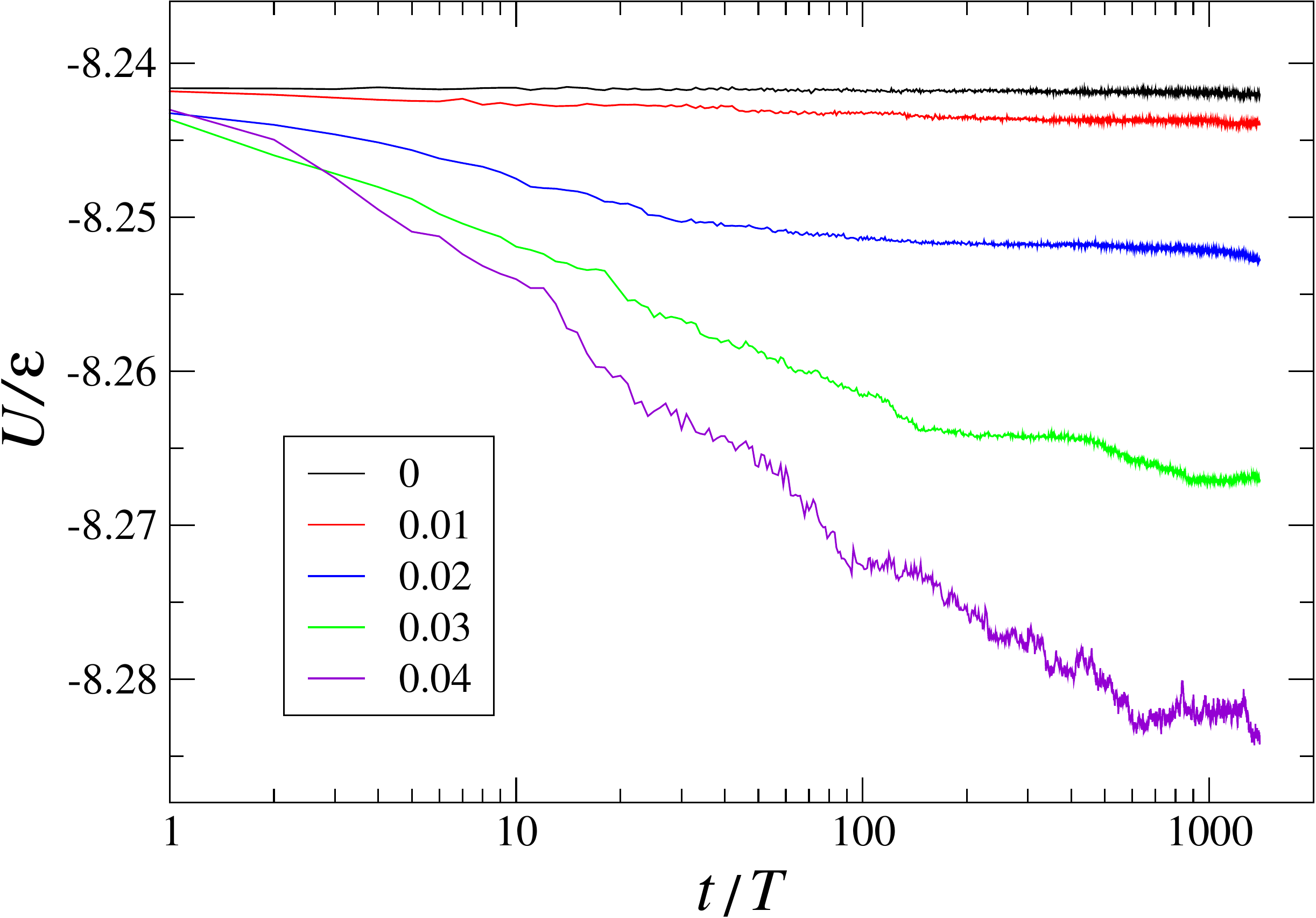}
\caption{(Color online) The potential energy minima at the end of
each cycle for the strain amplitudes $\varepsilon_{0}=0$ (black),
$0.01$ (red), $0.02$ (blue), $0.03$ (green), and $0.04$ (violet),
from top to bottom. The period of cyclic deformation is
$T=5000\,\tau$ and the system temperature is
$T_{LJ}=0.01\,\varepsilon/k_B$. The data for $\varepsilon_{0}=0$
correspond to the quiescent glass annealed during the time interval
$1400\,T$.}
\label{fig:poten_min_amp}
\end{figure}

The time dependence of the potential energy minima (at the end of
each cycle) is presented in Fig.\,\ref{fig:poten_min_amp} for the
strain amplitudes, $0\leqslant \varepsilon_0 \leqslant 0.04$, in the
elastic range.   The data for the undeformed glass (at
$\varepsilon_0=0$) represent the aging process at constant volume,
and it shows that the potential energy remains essentially constant
during the time interval of $1400\,T$. By contrast, it can be
clearly observed that periodic deformation at larger strain
amplitudes results in deeper energy minima. The lowest value of the
potential energy, $U\approx-8.284\,\varepsilon$, is attained at
$\varepsilon_0=0.04$ at $t\approx 1400\,T$. Interestingly, this
value is nearly the same as the potential energy level reported in
the previous study on periodic shear strain along a single plane at
the strain amplitude, $\gamma_0=0.065$, just below the yielding
strain~\cite{NVP19alt}. We also find that the strain amplitude
$\varepsilon_0=0.05$ is greater than the critical strain amplitude,
and the periodic deformation of the glassy material involves
extended plastic flow and shear band formation after about 20 cycles
(not shown).

\vskip 0.05in


It should be noted that the potential energy minima are plotted in
Fig.\,\ref{fig:poten_min_amp} starting from the first cycle,
\textit{i.e.}, $t=T$. The potential energy right after rapid cooling
but before cyclic loading is $U\approx -8.20\,\varepsilon$ (not
shown), and after the first cycle it is reduced to $U\approx
-8.24\,\varepsilon$, depending on the strain amplitude (see
Fig.\,\ref{fig:poten_min_amp}). In other words, upon rapid cooling,
the glass is settled at a relatively high energy state, and during
the first period (either at mechanical equilibrium,
$\varepsilon_0=0$, or periodically strained, $\varepsilon_0>0$) the
samples relax to $U\approx-8.24\,\varepsilon$ via large-scale
particle rearrangements, and only then a more gradual relaxation
process begins (as shown in Fig.\,\ref{fig:poten_min_amp}).
Therefore, in the following analysis of the relaxation dynamics, the
reference configuration of atoms is taken at $t=T$ for each value of
the strain amplitude.

\vskip 0.05in


The microscopic details of the structural relaxation dynamics during
cyclic loading can be inspected via the analysis of the so-called
nonaffine displacements of atoms. Recall that atomic displacements
in deformed crystalline solids is defined with respect to the
periodic lattice. By contrast, the local displacement of atoms in
disordered materials can be measured with respect to their
neighbors. In this case, the displacement consists of two parts,
\textit{i.e.}, affine and nonaffine components. In turn, the
nonaffine part can be computed numerically using the transformation
matrix $\mathbf{J}_i$, which linearly transforms a group of
neighbors around the atom $i$ during the time interval $\Delta t$
and, at the same time, minimizes the quantity:
\begin{equation}
D^2(t, \Delta t)=\frac{1}{N_i}\sum_{j=1}^{N_i}\Big\{
\mathbf{r}_{j}(t+\Delta t)-\mathbf{r}_{i}(t+\Delta t)-\mathbf{J}_i
\big[ \mathbf{r}_{j}(t) - \mathbf{r}_{i}(t)    \big] \Big\}^2,
\label{Eq:D2min}
\end{equation}
where the summation is carried over the neighboring atoms within a
sphere of radius $1.5\,\sigma$ located at $\mathbf{r}_{i}(t)$. In
the original study, Falk and Langer showed that the nonaffine
measure, given by Eq.\,(\ref{Eq:D2min}), is an excellent diagnostic
for detecting local shear transformations when the time interval
$\Delta t$ is properly chosen~\cite{Falk98}.  In the last several
years, the analysis on nonaffine displacements was applied to
amorphous materials subjected to time periodic
deformation~\cite{Priezjev16,Priezjev16a,Priezjev17,Priezjev18,
Priezjev18a,NVP18strload,NVP19alt}, elastostatic
loading~\cite{PriezELAST19}, and thermal
cycling~\cite{Priez19tcyc,Priez19T2000,Priez19T5000,Priez19one}.  In
particular, it was found that during periodic shear deformation at
strain amplitudes above the critical value, both well and poorly
annealed glasses undergo a yielding transition after a number of
transient cycles~\cite{Priezjev17,Priezjev18a}. Moreover, the
spatial distribution of nonaffine displacements changes from a set
of disconnected clusters of mobile atoms to a localized shear
band~\cite{Priezjev17,Priezjev18a,Priez19ba}.

\vskip 0.05in

\begin{figure}[t]
\includegraphics[width=12.cm,angle=0]{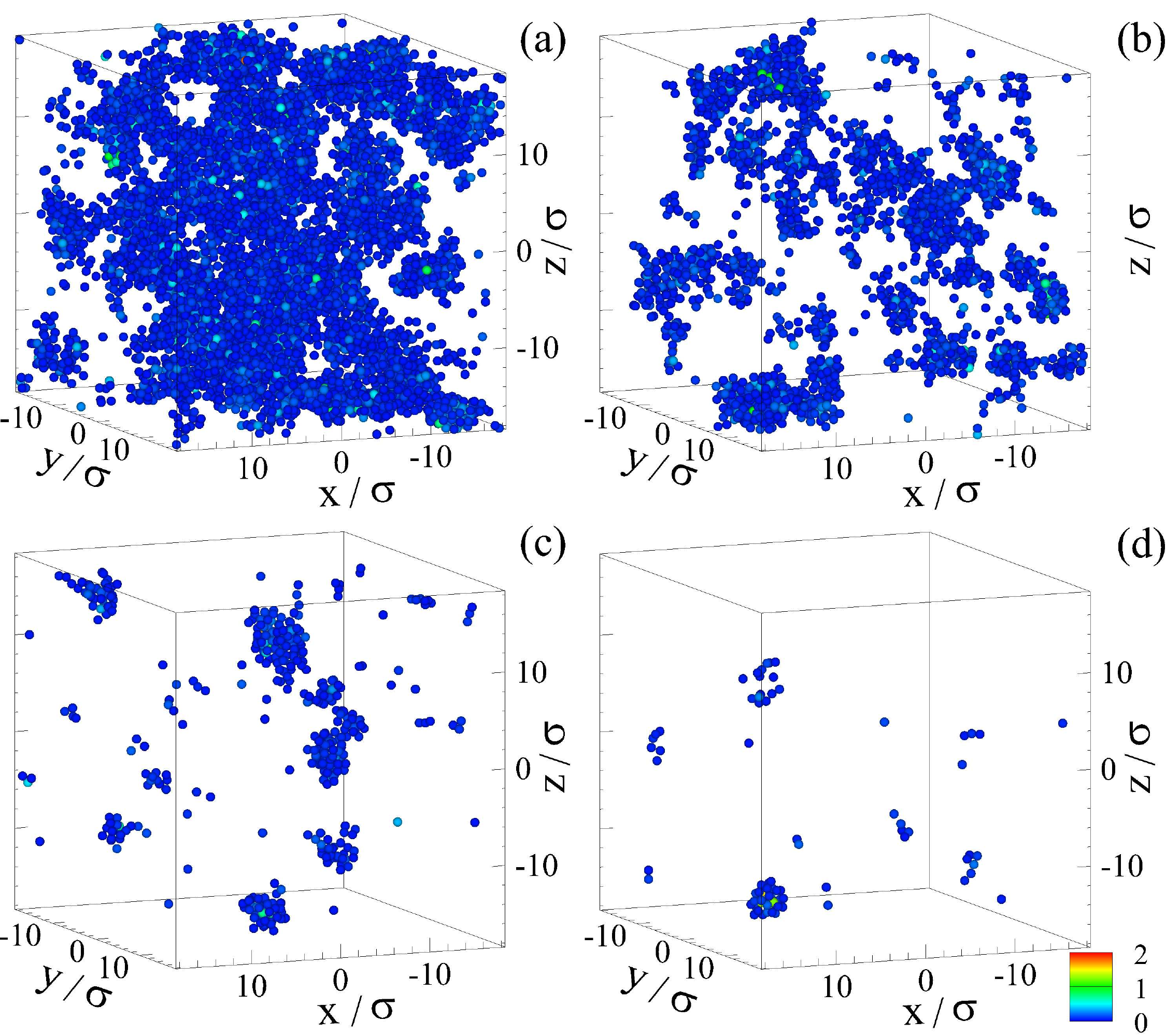}
\caption{(Color online) The positions of atoms with large nonaffine
measure (a) $D^2(T,T)>0.04\,\sigma^2$, (b)
$D^2(10T,T)>0.04\,\sigma^2$, (c) $D^2(100T,T)>0.04\,\sigma^2$, and
(d) $D^2(1000T,T)>0.04\,\sigma^2$. The color code for $D^2(nT,T)$ is
defined in the legend.  The strain amplitude is
$\varepsilon_{0}=0.03$ and the oscillation period is $T=5000\,\tau$.
}
\label{fig:snapshot_clusters_cyclic_003}
\end{figure}
\begin{figure}[t]
\includegraphics[width=12.cm,angle=0]{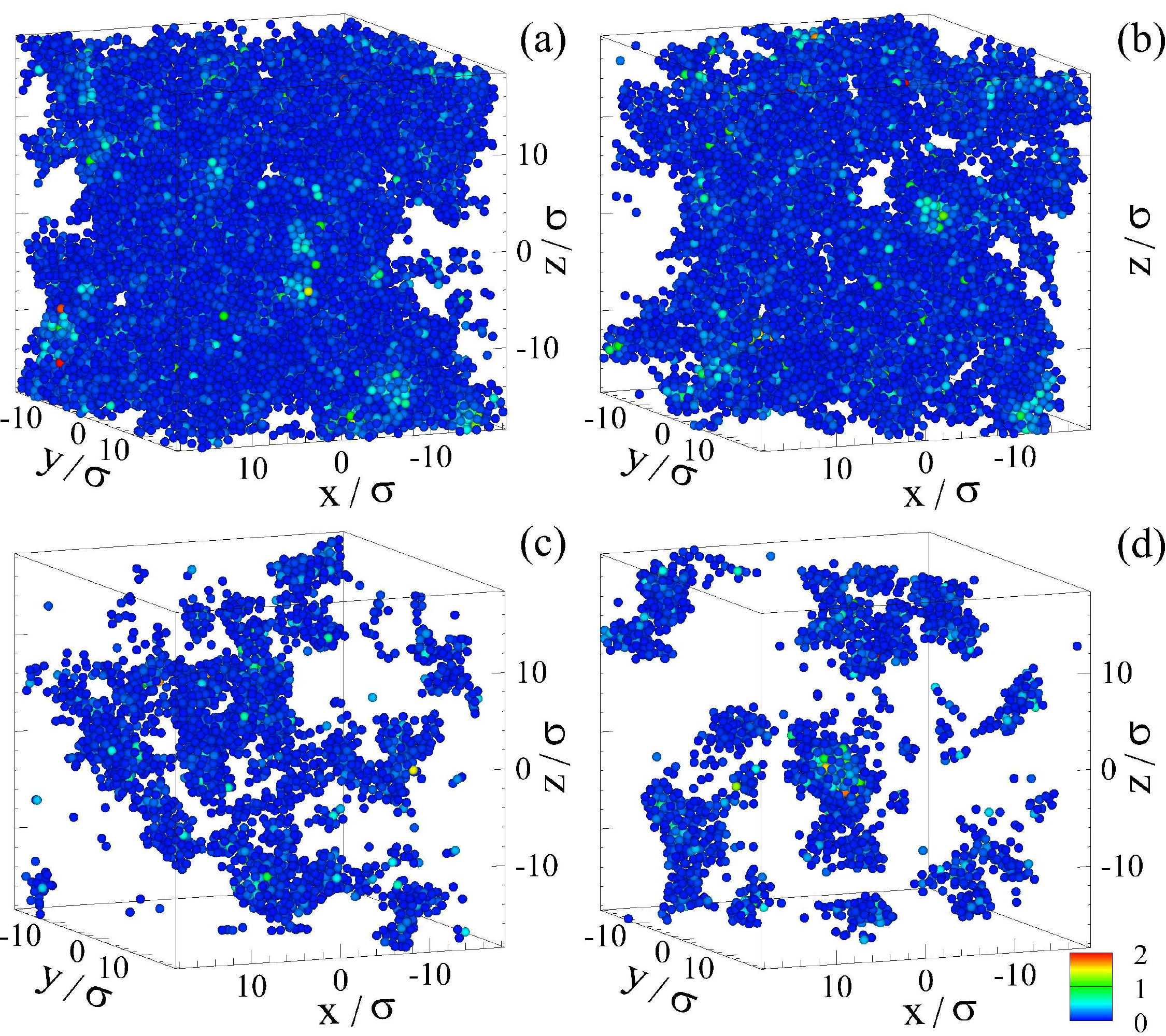}
\caption{(Color online) The sequence of snapshots of atoms with the
nonaffine measure (a) $D^2(T,T)>0.04\,\sigma^2$, (b)
$D^2(10T,T)>0.04\,\sigma^2$, (c) $D^2(100T,T)>0.04\,\sigma^2$, and
(d) $D^2(1000T,T)>0.04\,\sigma^2$. The atoms are colored according
to the legend. The oscillation period is $T=5000\,\tau$ and the
strain amplitude is $\varepsilon_{0}=0.04$. }
\label{fig:snapshot_clusters_cyclic_004}
\end{figure}

The data for the potential energy reported in
Fig.\,\ref{fig:poten_min_amp} indicate that the relaxation dynamics
for each value of the strain amplitude slows down as the cycle
number increases. This, in turn, implies that the typical size of
regions, where atoms undergo irreversible displacements to lower
energy states, should also be reduced over time. The sequence of
snapshots of atoms with relatively large nonaffine displacements,
$D^2(nT,T)>0.04\,\sigma^2$, are shown in
Fig.\,\ref{fig:snapshot_clusters_cyclic_003} for the strain
amplitude $\varepsilon_0=0.03$ and in
Fig.\,\ref{fig:snapshot_clusters_cyclic_004} for
$\varepsilon_0=0.04$.  Here, the nonaffine displacements,
Eq.\,(\ref{Eq:D2min}), were computed for pairs of configurations at
zero strain separated by the time interval $\Delta t = T$, where
$T=5000\,\tau$ is the oscillation period. It can be clearly observed
in Figs.\,\ref{fig:snapshot_clusters_cyclic_003} and
\ref{fig:snapshot_clusters_cyclic_004} that the initial stage of
relaxation involves percolating clusters of mobile atoms, and their
typical size is gradually reduced as the cycle number increases.
Note that after 1000 cycles at $\varepsilon_0=0.03$, shown in
Fig.\,\ref{fig:snapshot_clusters_cyclic_003}\,(d), almost all atoms
remained inside cages formed by their neighbors during one cycle.
For reference, the cage size at $\rho=1.2\,\sigma^{-3}$ is about
$0.1\,\sigma$~\cite{KobAnd95,Priezjev16,Priezjev16a}. This behavior
is consistent with the results of previous MD studies on periodic
shear deformation of poorly annealed binary glasses, where it was
shown that the volume occupied by atoms with large nonaffine
displacements is decreased over time if the strain amplitude is
below the critical value~\cite{Priezjev18,Priezjev18a}.

\vskip 0.05in

\begin{figure}[t]
\includegraphics[width=10.0cm,angle=0]{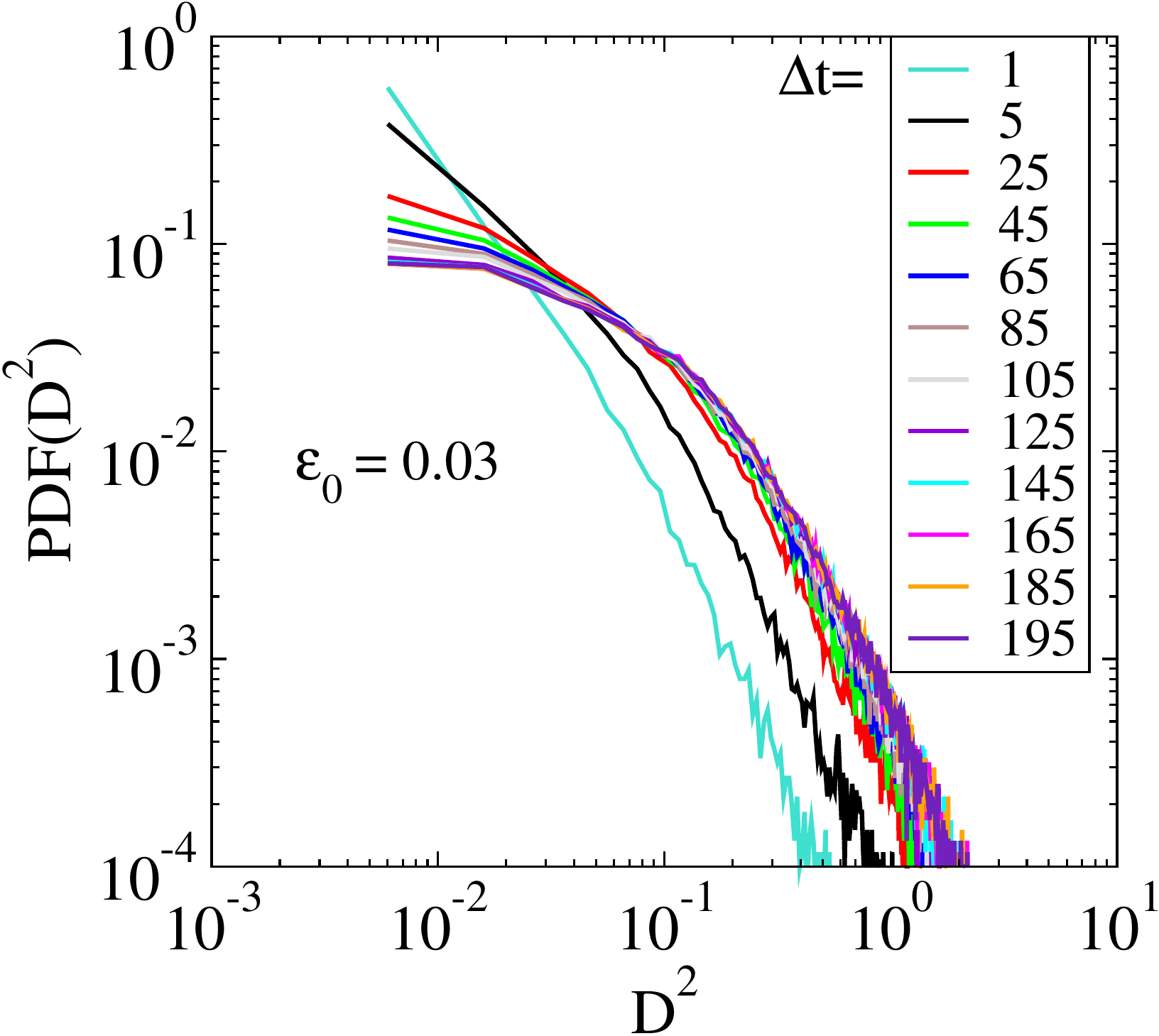}
\caption{(Color online) The distribution of nonaffine quantity
$D^2(t_1,\Delta t)$ for the strain amplitude $\varepsilon_{0}=0.03$.
The legend shows time intervals $\Delta t$, measured in oscillation
periods ($T=5000\,\tau$), with respect to the atomic configuration
at $t_1=T$. }
\label{fig:PD2_eps003}
\end{figure}
\begin{figure}[t]
\includegraphics[width=10.0cm,angle=0]{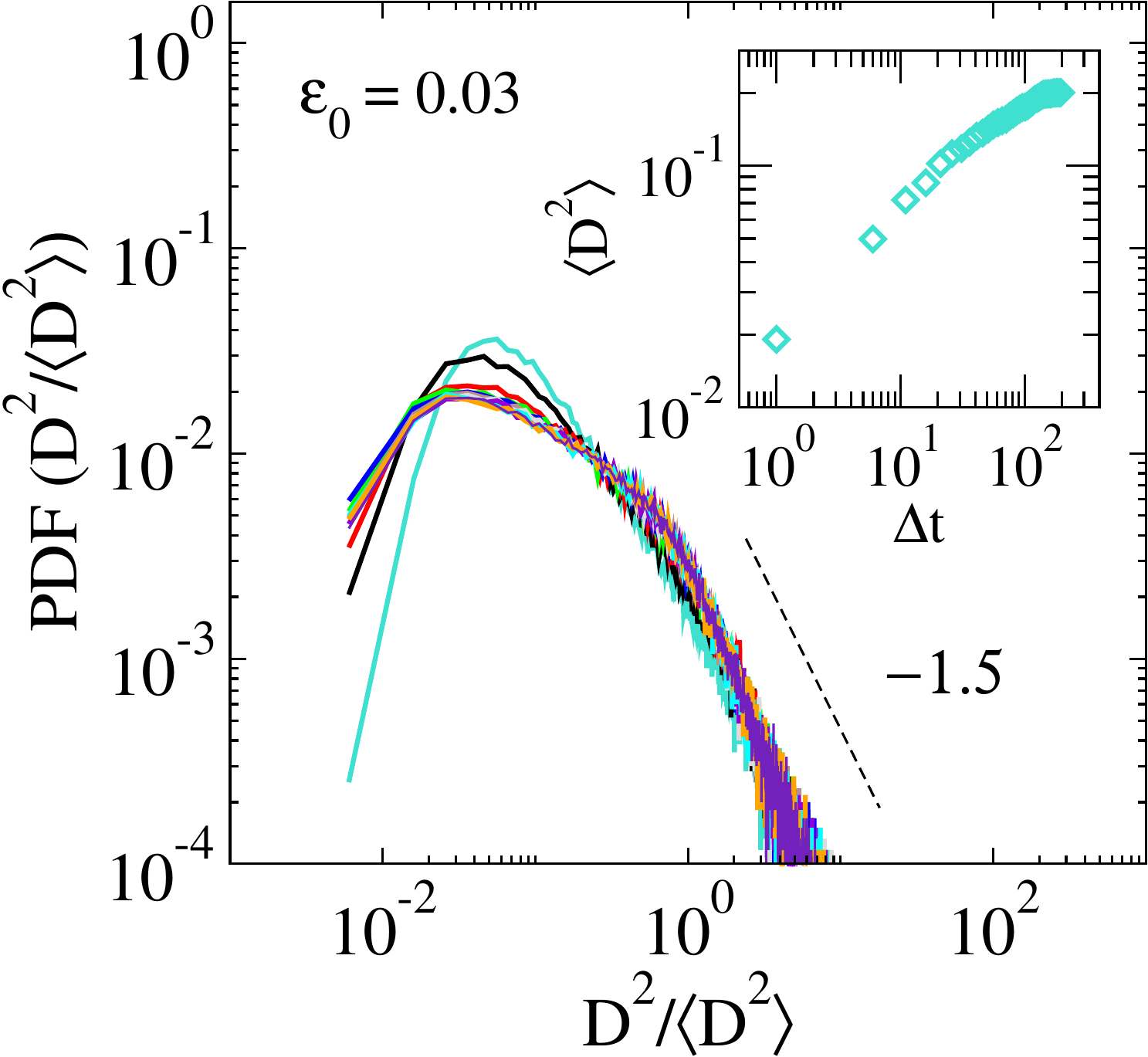}
\caption{(Color online) The rescaled distribution of $D^2(t_1,\Delta
t)$ for the strain amplitude $\varepsilon_{0}=0.03$. The same data
and color code as in Fig.\,\ref{fig:PD2_eps003}. The dashed line
indicates the slope $-1.5$.    The inset shows the average of
$D^2(t_1,\Delta t)$ as a function of the time interval $\Delta t$.}
\label{fig:PD2_re_eps003}
\end{figure}
\begin{figure}[t]
\includegraphics[width=10.0cm,angle=0]{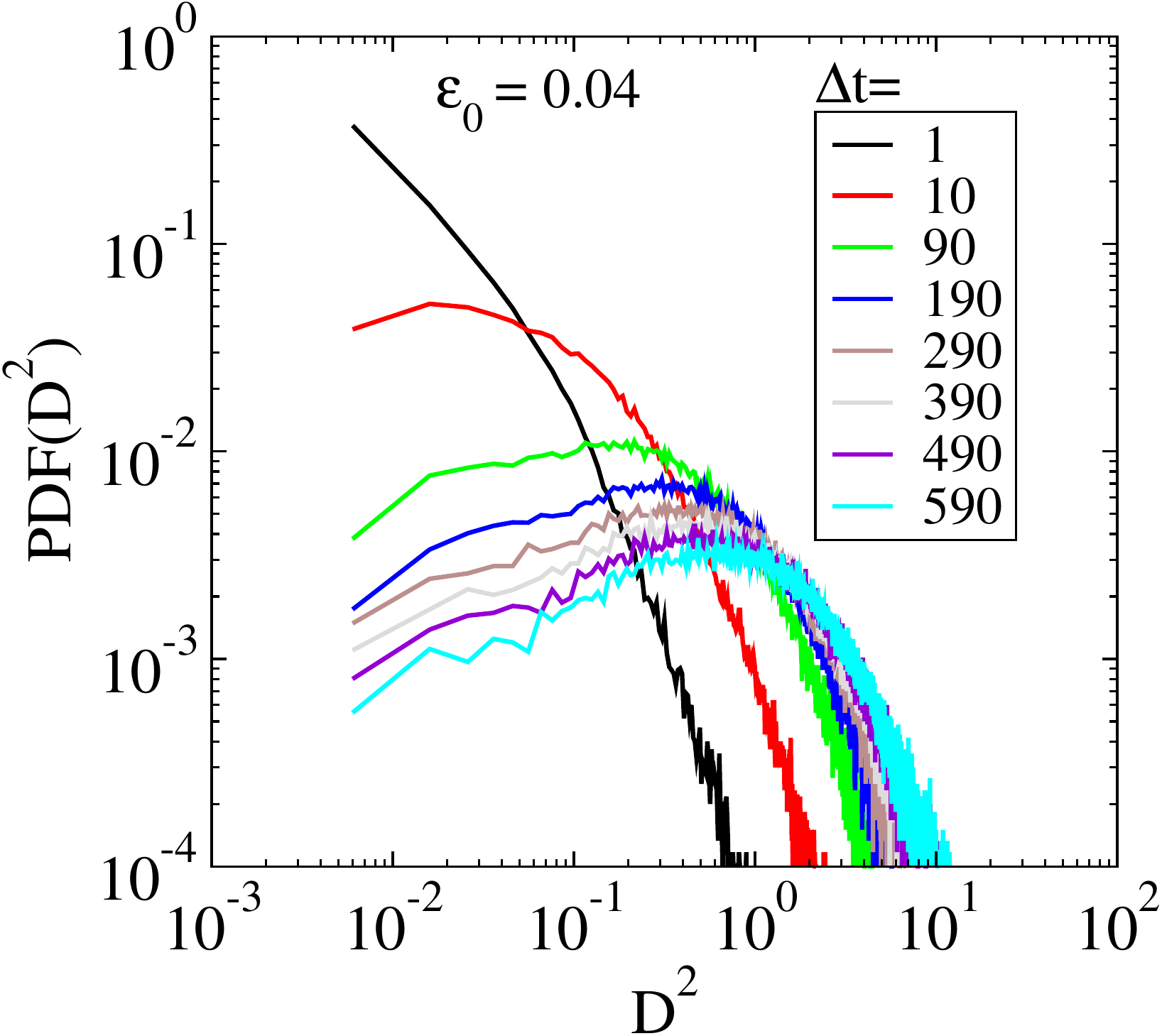}
\caption{(Color online) The probability distribution of the
nonaffine measure $D^2(t_1,\Delta t)$ for the strain amplitude
$\varepsilon_{0}=0.04$.  The values of the time interval $\Delta t$
are listed in the table. The reference state is $t_1=T$, where
$T=5000\,\tau$ is the period of cyclic loading. }
\label{fig:PD2_eps004}
\end{figure}
\begin{figure}[t]
\includegraphics[width=10.0cm,angle=0]{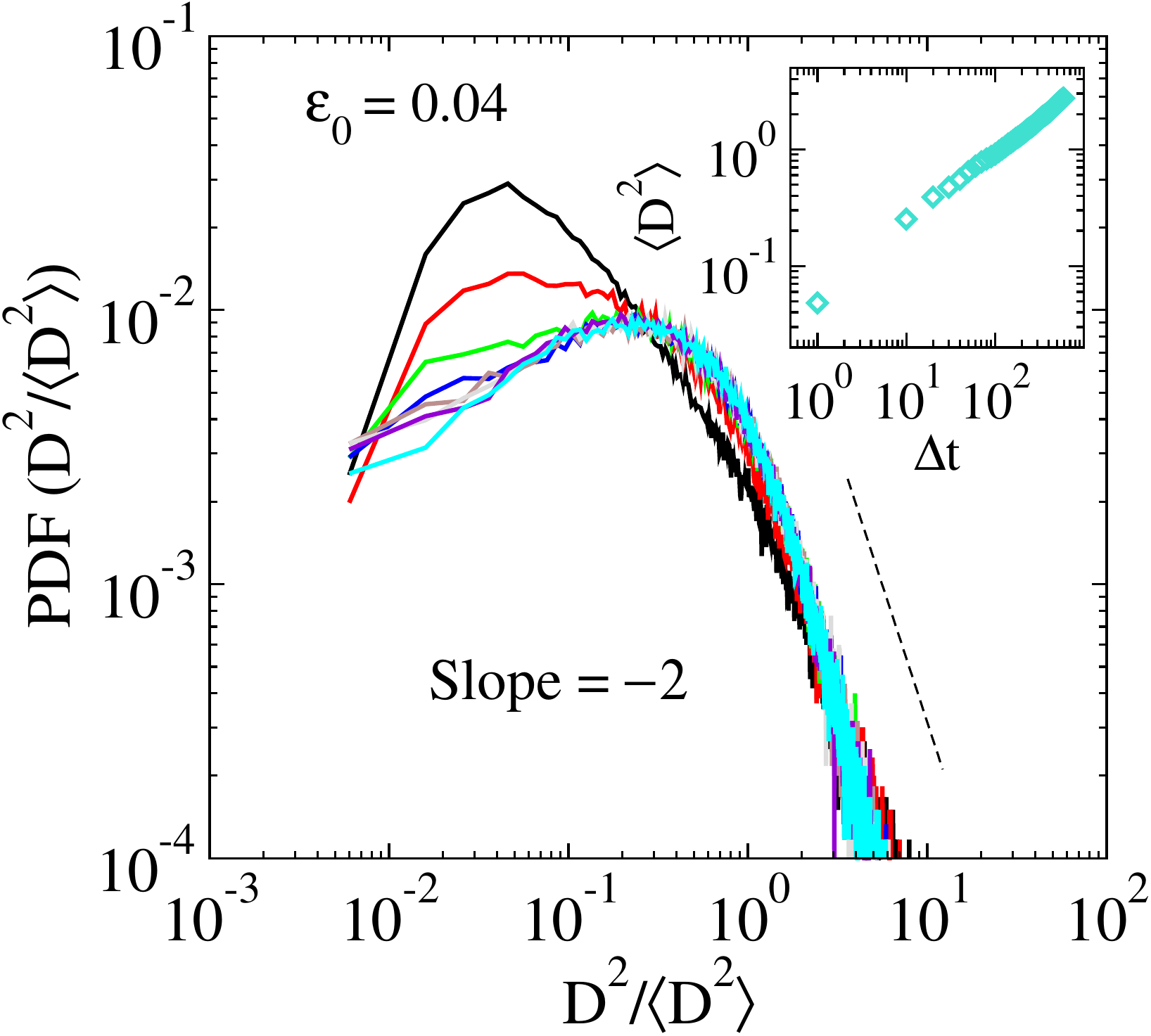}
\caption{(Color online) The distribution of $D^2(t_1,\Delta t)$
divided by its average for the strain amplitude
$\varepsilon_{0}=0.04$. The color code is the same as in
Fig.\,\ref{fig:PD2_eps004}. The slope $-2$ is shown for reference.
The average value of $D^2(t_1,\Delta t)$ versus $\Delta t$ is
reported in the inset. }
\label{fig:PD2_re_eps004}
\end{figure}

The probability distribution functions of nonaffine displacements
are plotted in Figs.\,\ref{fig:PD2_eps003} and
\ref{fig:PD2_re_eps003} for the strain amplitude
$\varepsilon_0=0.03$ and in Figs.\,\ref{fig:PD2_eps004} and
\ref{fig:PD2_re_eps004} for $\varepsilon_0=0.04$. In both cases, the
nonaffine displacements were computed with respect to the
configurations at $t_1=T$ during time intervals when the decay of
the potential energy in Fig.\,\ref{fig:poten_min_amp} is roughly
linear (on the log-normal scale). More specifically, the time
interval is $\Delta t \lesssim 200\,T$ for $\varepsilon_0=0.03$ and
$\Delta t \lesssim 600\,T$ for $\varepsilon_0=0.04$.  It can be seen
in Figs.\,\ref{fig:PD2_eps003}\,(a) and \ref{fig:PD2_eps004}\,(a)
that the distributions become relatively broad already at $\Delta t
= T$, which correlates well with the appearance of large clusters in
Figs.\,\ref{fig:snapshot_clusters_cyclic_003}\,(a) and
\ref{fig:snapshot_clusters_cyclic_004}\,(a). As expected, the width
of the distributions increases at larger time intervals (see
Figs.\,\ref{fig:PD2_eps003} and \ref{fig:PD2_eps004}), which
reflects larger displacements of atoms with respect to their
neighbors at time $t_1=T$.   Interestingly, the probability
distribution of the nonaffine measure divided by its averaged value
gradually evolves towards a function that does not depend on the
cycle number, as shown in Figs.\,\ref{fig:PD2_re_eps003} and
\ref{fig:PD2_re_eps004}. The common shape of the distribution
functions of $D^2/\langle D^2\rangle$ is most probably related to a
weak correlation between atomic configurations separated by a large
time interval $\Delta t$.  In other words, most of the atoms undergo
multiple cage jumps, and the neighbors of an atom at $t_1=T$ are
displaced independently over sufficiently large $\Delta t$. Note,
however, that the slopes of the power-law decay are slightly
different for $\varepsilon_0=0.03$ and $0.04$.

\vskip 0.05in

\begin{figure}[t]
\includegraphics[width=12.0cm,angle=0]{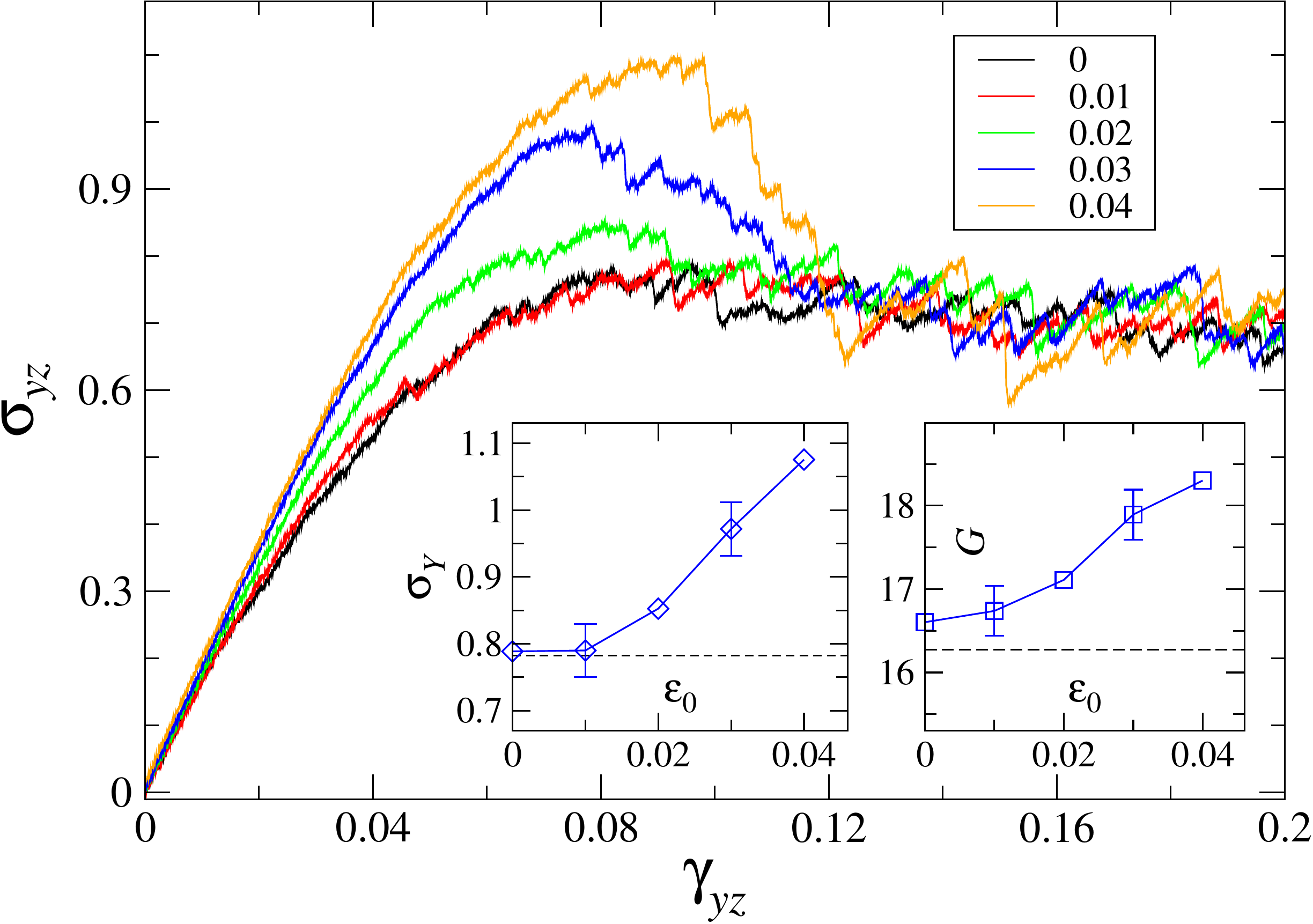}
\caption{(Color online) The variation of shear stress, $\sigma_{yz}$
(in units of $\varepsilon\sigma^{-3}$) during steady strain with the
rate $10^{-5}\,\tau^{-1}$.  The glasses were strained after 1400
tension-compression cycles with the indicated strain amplitudes. The
insets show the yielding peak, $\sigma_{Y}$ (in units of
$\varepsilon\sigma^{-3}$), and the shear modulus, $G$ (in units of
$\varepsilon\sigma^{-3}$), as functions of the strain amplitude. The
horizontal dashed lines in the insets indicate $\sigma_{Y}$ and $G$
before periodic loading was applied. }
\label{fig:stress_Y_G_vs_gam0}
\end{figure}

We next discuss the effect of cyclic loading at different strain
amplitudes on the mechanical properties of binary glasses. As shown
in Fig.\,\ref{fig:poten_min_amp}, upon increasing strain amplitude
within the elastic range, the systems are gradually relocated to
lower energy states. After 1400 cycles, the glass was subjected to
startup shear deformation along the $yz$ plane with the
computationally slow rate $\dot{\gamma}_{yz}=10^{-5}\,\tau^{-1}$.
The shear stress-strain curves are reported in
Fig.\,\ref{fig:stress_Y_G_vs_gam0} for glasses that were annealed
during $1400\,T$ and loaded during 1400 cycles at
$T_{LJ}=0.01\,\varepsilon/k_B$.  It can be observed that with
increasing strain amplitude of the cyclic loading, $\varepsilon_0$,
the initial slope of the shear stress increases and a pronounced
yielding peak is developed at the shear strain $\gamma_{yz}\approx
0.09$. These trends are consistent with the dependence of the
mechanical response of disordered solids on the degree of
relaxation; namely, more relaxed glasses are stronger and more
brittle~\cite{Greer16}.

\vskip 0.05in

\begin{figure}[t]
\includegraphics[width=12.cm,angle=0]{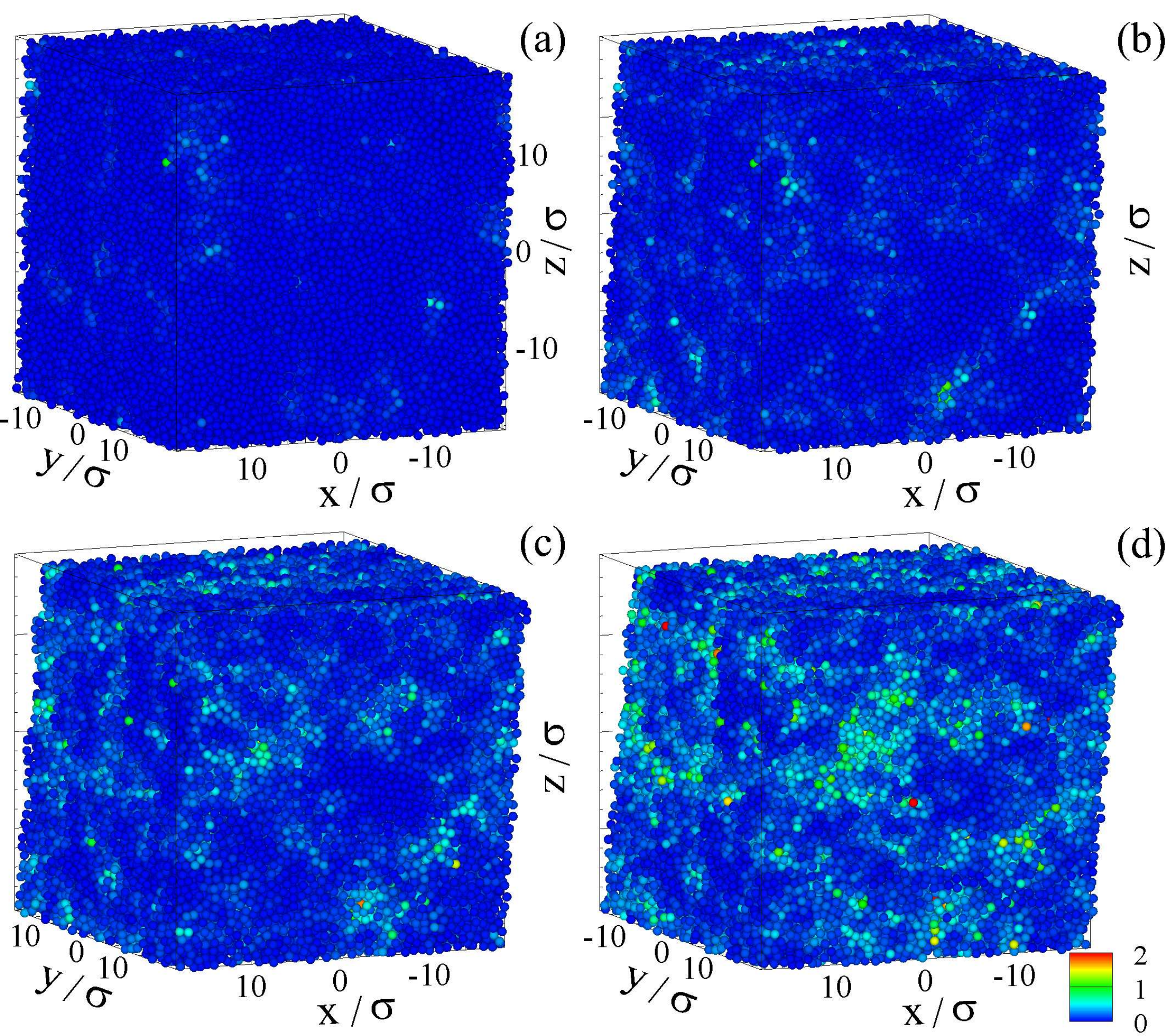}
\caption{(Color online) The snapshots of the binary glass that was
annealed during $1400\,T$ at $T_{LJ}=0.01\,\varepsilon/k_B$ and then
sheared along the $yz$ plane with the rate of $10^{-5}\,\tau^{-1}$.
The shear strain is (a) $0.05$, (b) $0.10$, (c) $0.15$, and (d)
$0.20$. The color denotes $D^2$ with respect to the atomic
configuration at zero strain.}
\label{fig:snap_amp00_Gyz}
\end{figure}
\begin{figure}[t]
\includegraphics[width=12.cm,angle=0]{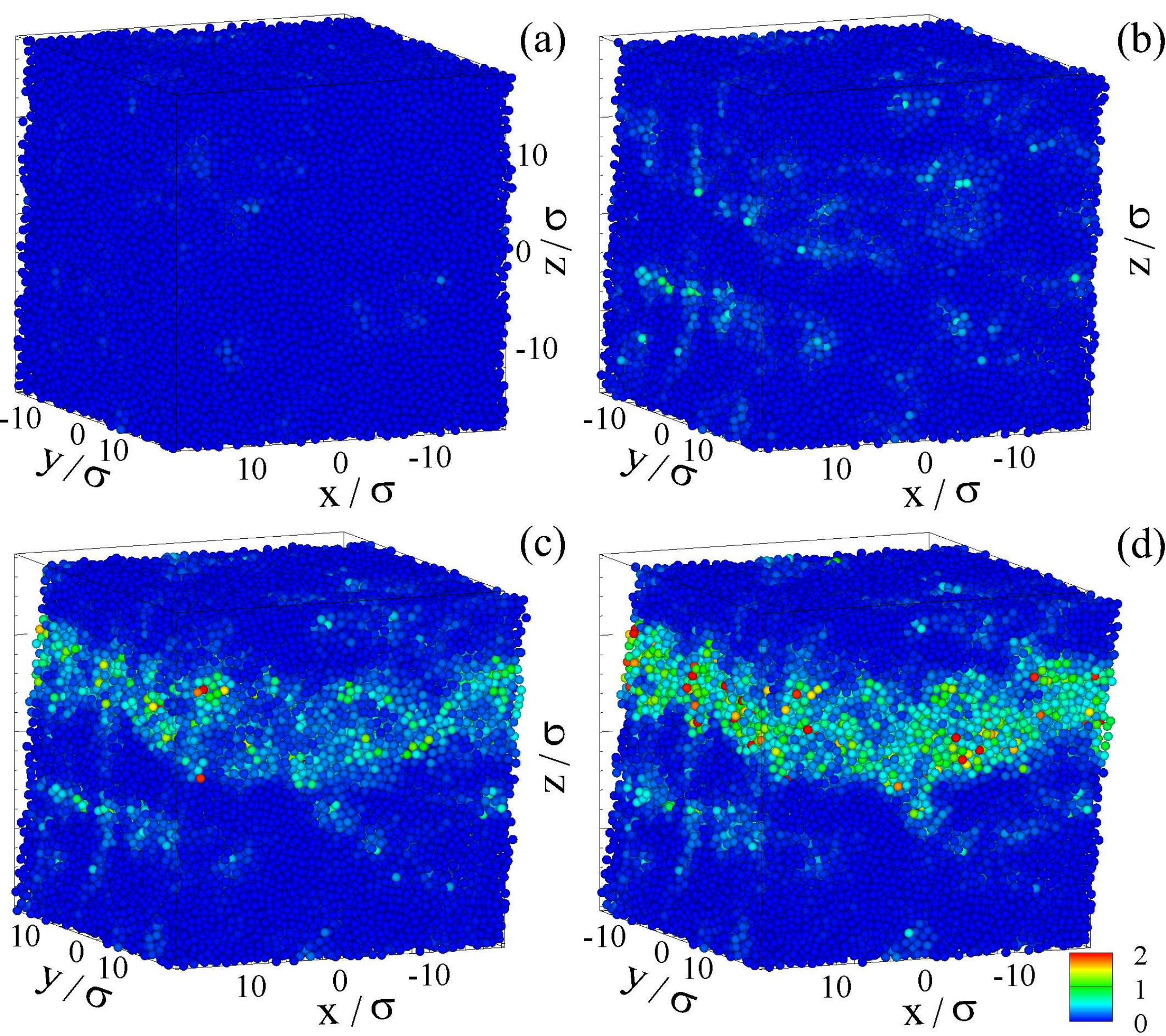}
\caption{(Color online) The atomic configurations of the strained
glass after $1400$ tension-compression cycles with the strain
amplitude $\varepsilon_0=0.04$. The shear strain, $\gamma_{yz}$, is
(a) $0.05$, (b) $0.10$, (c) $0.15$, and (d) $0.20$. The color of
atoms indicates their nonaffine displacements with respect to
$\gamma_{yz}=0$, see the legend. }
\label{fig:snap_amp04_Gyz}
\end{figure}

The dependence of the yielding peak, $\sigma_Y$, and shear modulus,
$G$, as functions of the strain amplitude is summarized in the
insets of Fig.\,\ref{fig:stress_Y_G_vs_gam0}. Here, the shear
modulus was computed from the slope of the linear fit of the
stress-strain curves at $\gamma_{yz}\leqslant0.01$, and the data
were averaged over three mutually perpendicular planes of shear. In
the insets, the dashed lines indicate the averaged values of
$\sigma_Y$ and $G$ before the cyclic loading was applied, whereas
the data for $\varepsilon_0=0$ were taken in quiescent samples
annealed during the time interval $1400\,T$. It can be clearly seen
that both the stress overshoot and the shear modulus increase at
larger strain amplitudes, which is in agreement with the trend
reported for the potential energy in Fig.\,\ref{fig:poten_min_amp}.
Furthermore, the spatial distributions of nonaffine displacements
during startup shear deformation are shown for the annealed sample
in Fig.\,\ref{fig:snap_amp00_Gyz} and for the periodically deformed
glass in Fig.\,\ref{fig:snap_amp04_Gyz}.  As is evident, the
processing history results in a distinct difference in the
deformation pattern, \textit{i.e.}, a nearly homogeneous deformation
of rapidly quenched and then thermally annealed sample, and the
appearance of the shear band in the cyclicly annealed glass.

\section{Conclusions}

In summary, we carried out molecular dynamics simulations to study
the relaxation dynamics in disordered solids subjected to
tension-compression cyclic loading and its effect on mechanical
properties. The model glass was represented via a binary mixture
with strongly non-additive cross interactions that prevent
crystallization upon cooling. The binary mixture was rapidly
quenched from the liquid phase to the glass phase at a temperature
well below the glass transition temperature. The deformation
protocol included periodic tension-compression deformation at
constant volume during hundreds of cycles. During cyclic loading,
the amorphous systems were driven to progressively lower potential
energy states via collective nonaffine rearrangements of atoms.
Moreover, it was shown that the shape of the rescaled probability
distribution function becomes time independent after sufficiently
large number of cycles. After cyclic loading, the mechanical
properties were probed during startup shear deformation at a
constant strain rate. The results of numerical simulations indicate
that both the shear modulus and the peak value of the stress
overshoot increase in glasses that were periodically loaded with
larger strain amplitudes within the elastic range.

\section*{Acknowledgments}

Funding from the National Science Foundation (CNS-1531923) is
gratefully acknowledged. The molecular dynamics simulations were
performed using the LAMMPS open-source code developed at Sandia
National Laboratories~\cite{Lammps}. The simulations were performed
at Wright State University's Computing Facility and the Ohio
Supercomputer Center.


%

%

%

%

%

%

%

%
%

%
%

%
%

\bibliographystyle{prsty}

\begin{thebibliography}{99}


\bibitem{Khan18}       M.~M. Khan, A. Nemati, Z.~U. Rahman, U.~H. Shah, H. Asgar, and W. Haider,
                       Recent Advancements in Bulk Metallic Glasses and Their Applications: A Review,
                       Crit. Rev. Solid State Mater. Sci. {\bf 43}, 233 (2018).


\bibitem{Spaepen77}    F. Spaepen,
                       A microscopic mechanism for steady state inhomogeneous flow in metallic glasses,
                       Acta Metall. {\bf 25}, 407 (1977).


\bibitem{Argon79}      A.~S. Argon,
                       Plastic deformation in metallic glasses,
                       Acta Metall. {\bf 27}, 47 (1979).

\bibitem{Egami13}      T. Egami, T. Iwashita, and W. Dmowski,
                       Mechanical properties of metallic glasses,
                       Metals {\bf 3}, 77 (2013).


\bibitem{Greer16}      Y. Sun, A. Concustell, and A.~L. Greer,
                       Thermomechanical processing of metallic glasses: Extending the range of the glassy state,
                       Nat. Rev. Mater. {\bf 1}, 16039 (2016).




\bibitem{Priezjev13}   N.~V. Priezjev,
                       Heterogeneous relaxation dynamics in amorphous materials under cyclic loading,
                       Phys. Rev. E {\bf 87}, 052302 (2013).

\bibitem{Sastry13}     D. Fiocco, G. Foffi, and S. Sastry,
                       Oscillatory athermal quasistatic deformation of a model glass,
                       Phys. Rev. E {\bf 88}, 020301(R) (2013).

\bibitem{Reichhardt13} I. Regev, T. Lookman, and C. Reichhardt,
                       Onset of irreversibility and chaos in amorphous solids under periodic shear,
                       Phys. Rev. E {\bf 88}, 062401 (2013).

\bibitem{Priezjev14}   N.~V. Priezjev,
                       Dynamical heterogeneity in periodically deformed polymer glasses,
                       Phys. Rev. E {\bf 89}, 012601 (2014).


\bibitem{Shi15}        J. Luo, K. Dahmen, P.~K. Liaw, and Y. Shi,
                       Low-cycle fatigue of metallic glass nanowires,
                       Acta Mater. {\bf 87}, 225 (2015).

\bibitem{Yang16}       Y.~F. Ye, S. Wang, J. Fan, C.~T. Liu, and Y. Yang,
                       Atomistic mechanism of elastic softening in metallic glass under cyclic
                       loading revealed by molecular dynamics simulations,
                       Intermetallics {\bf 68}, 5 (2016).

\bibitem{Priezjev16}   N.~V. Priezjev,
                       Reversible plastic events during oscillatory deformation of amorphous solids,
                       Phys. Rev. E {\bf 93}, 013001 (2016).

\bibitem{Kawasaki16}   T. Kawasaki and L. Berthier,
                       Macroscopic yielding in jammed solids is accompanied by a non-equilibrium
                       first-order transition in particle trajectories,
                       Phys. Rev. E {\bf 94}, 022615 (2016).

\bibitem{Priezjev16a}  N.~V. Priezjev,
                       Nonaffine rearrangements of atoms in deformed and quiescent binary glasses,
                       Phys. Rev. E {\bf 94}, 023004 (2016).

\bibitem{Sastry17}     P. Leishangthem, A.~D.~S. Parmar, and S. Sastry,
                       The yielding transition in amorphous solids under oscillatory shear deformation,
                       Nat. Commun. {\bf 8}, 14653 (2017).

\bibitem{Priezjev17}   N.~V. Priezjev,
                       Collective nonaffine displacements in amorphous materials during large-amplitude oscillatory shear,
                       Phys. Rev. E {\bf 95}, 023002 (2017).

\bibitem{Zapperi17}    P.~K. Jana, M.~J. Alava, and S. Zapperi,
                       Irreversibility transition of colloidal polycrystals under cyclic deformation,
                       Sci. Rep. {\bf 7}, 45550 (2017).

\bibitem{Hecke17}      S. Dagois-Bohy, E. Somfai, B.~P. Tighe, and M. van Hecke,
                       Softening and yielding of soft glassy materials,
                       Soft Matter {\bf 13}, 9036 (2017).

\bibitem{Priezjev18}   N.~V. Priezjev,
                       Molecular dynamics simulations of the mechanical annealing process in
                       metallic glasses: Effects of strain amplitude and temperature,
                       J. Non-Cryst. Solids {\bf 479}, 42 (2018).

\bibitem{Alava18}      P.~K. Jana, M.~J. Alava, and S. Zapperi,
                       Irreversible transition of amorphous and polycrystalline colloidal solids under cyclic deformation,
                       Phys. Rev. E {\bf 98}, 062607 (2018).

\bibitem{Priezjev18a}  N.~V. Priezjev,
                       The yielding transition in periodically sheared binary glasses at finite temperature,
                       Comput. Mater. Sci. {\bf 150}, 162 (2018).

\bibitem{NVP18strload} N.~V. Priezjev,
                       Slow relaxation dynamics in binary glasses during stress-controlled,
                       tension-compression cyclic loading,
                       Comput. Mater. Sci. {\bf 153}, 235 (2018).

\bibitem{She19}        Y. Bai and C. She,
                       Atomic structure evolution in metallic glasses under cyclic deformation,
                       Comput. Mater. Sci. {\bf 169}, 109094 (2019).

\bibitem{PriMakrho05}  N.~V. Priezjev and M.~A. Makeev,
                       The influence of periodic shear on structural relaxation and pore redistribution in binary glasses,
                       J. Non-Cryst. Solids {\bf 506}, 14 (2019).

\bibitem{PriMakrho09}  N.~V. Priezjev and M.~A. Makeev,
                       Structural transformations during periodic deformation of low-porosity amorphous materials,
                       Modelling Simul. Mater. Sci. Eng. {\bf 27}, 025004 (2019).

\bibitem{Deng19}       M. Zhang, Q. Li, J. Zhang, X. Wang, J. Jin, P. Gong, and L. Deng,
                       Influence of vibrational loading on deformation
                       behavior of metallic glass: A molecular dynamics study,
                       Metals {\bf 9}, 1197 (2019).

\bibitem{NVP19alt}     N.~V. Priezjev,
                       Accelerated relaxation in disordered solids under cyclic loading with alternating shear orientation,
                       J. Non-Cryst. Solids {\bf 525}, 119683 (2019).

\bibitem{Priez19ba}    N.~V. Priezjev,
                       Shear band formation in amorphous materials under oscillatory shear deformation,
                       arXiv:1911.06157







\bibitem{KobAnd95}     W. Kob and H.~C. Andersen,
                       Testing mode-coupling theory for a supercooled binary Lennard-Jones mixture:
                       The van Hove correlation function,
                       Phys. Rev. E {\bf 51}, 4626 (1995).


\bibitem{Lammps}       S.~J. Plimpton,
                       Fast parallel algorithms for short-range molecular dynamics,
                       J. Comp. Phys. {\bf 117}, 1 (1995).


\bibitem{Allen87}      M.~P. Allen and D.~J. Tildesley,
                       {\it Computer Simulation of Liquids} (Clarendon, Oxford, 1987).

\bibitem{Stillinger00} M. Utz, P.~G. Debenedetti, and F.~H. Stillinger,
                       Atomistic simulation of aging and rejuvenation in glasses,
                       Phys. Rev. Lett. {\bf 84}, 1471 (2000).

\bibitem{Lacks04}      D.~J. Lacks and M.~J. Osborne,
                       Energy landscape picture of overaging and rejuvenation in a sheared glass,
                       Phys. Rev. Lett. {\bf 93}, 255501 (2004).

\bibitem{Falk98}       M.~L. Falk and J.~S. Langer,
                       Dynamics of viscoplastic deformation in amorphous solids,
                       Phys. Rev. E {\bf 57}, 7192 (1998).

\bibitem{Priez19tcyc}  N.~V. Priezjev,
                       The effect of cryogenic thermal cycling on aging, rejuvenation,
                       and mechanical properties of metallic glasses,
                       J. Non-Cryst. Solids {\bf 503}, 131 (2019).

\bibitem{Priez19T2000} Q.-L. Liu and N.~V. Priezjev,
                       The influence of complex thermal treatment on mechanical properties of amorphous materials,
                       Comput. Mater. Sci. {\bf 161}, 93 (2019).

\bibitem{Priez19T5000} N.~V. Priezjev,
                       Potential energy states and mechanical properties of thermally cycled binary glasses,
                       J. Mater. Res. {\bf 34}, 2664 (2019).

\bibitem{Priez19one}   N.~V. Priezjev,
                       Atomistic modeling of heat treatment processes for tuning the mechanical properties of disordered solids,
                       J. Non-Cryst. Solids {\bf 518}, 128 (2019).

\bibitem{PriezELAST19} N.~V. Priezjev,
                       Aging and rejuvenation during elastostatic loading of amorphous alloys:
                       A molecular dynamics simulation study,
                       Comput. Mater. Sci. {\bf 168}, 125 (2019).


\end{thebibliography}

\end{document}